\documentclass[aps,prd,twocolumn,superscriptaddress]{revtex4}
\usepackage{epsfig,epsf}
\usepackage{amsmath}
\usepackage{amsthm}
\usepackage{amsfonts}
\usepackage{amssymb}
\usepackage{dsfont}
\usepackage{multirow}
\usepackage{appendix}
\usepackage{slashed}
\usepackage[active]{srcltx}
\usepackage{psfrag}

\begin{document}

\title{New scalar resonance $X_0(2900)$ as a $\overline{D}^{*}K^{*}$
molecule: Mass and width}
\date{\today}
\author{S.~S.~Agaev}
\affiliation{Institute for Physical Problems, Baku State University, Az--1148 Baku,
Azerbaijan}
\author{K.~Azizi}
\affiliation{Department of Physics, University of Tehran, North Karegar Avenue, Tehran
14395-547, Iran}
\affiliation{Department of Physics, Do\v{g}u\c{s} University, Acibadem-Kadik\"{o}y, 34722
Istanbul, Turkey}
\affiliation{School of Particles and Accelerators, Institute for Research in Fundamental
Sciences (IPM) P.O. Box 19395-5531, Tehran, Iran}
\author{H.~Sundu}
\affiliation{Department of Physics, Kocaeli University, 41380 Izmit, Turkey}

\begin{abstract}
We explore features of the scalar structure $X_0(2900)$, which is one of the
two resonances discovered recently by LHCb in the $D^{-}K^{+}$ invariant
mass distribution in the decay $B^{+} \to D^{+}D^{-}K^{+}$. We treat $%
X_0(2900)$ as a hadronic molecule composed of the conventional mesons $%
\overline{D}^{* 0} $ and $K^{*0}$ and calculate its mass, coupling and
width. The mass and coupling of $X_0(2900)$ are determined using the QCD
two-point sum rule method by taking into account quark, gluon, and mixing
vacuum condensates up to dimension $15$. The decay of this structure to
final state $D^{-}K^{+}$ is investigated in the context of the light-cone
sum rule approach supported by a soft-meson technique. To this end, we
evaluate strong coupling $G$ corresponding to vertex $X_0D^{-}K^{+}$, which
allows us to find width of the decay $X_0(2900) \to D^{-}K^{+}$. Obtained
predictions for the mass of the hadronic molecule $\overline{D}^{*0 }K^{*0}$
$m=(2868 \pm 198)~\mathrm{MeV} $ and for its width $\Gamma=(49.6 \pm 9.3)~%
\mathrm{MeV}$ can be considered as arguments in favor of molecule
interpretation of $X_0(2900)$.
\end{abstract}

\maketitle


\section{Introduction}

\label{sec:Int} 

The LHCb collaboration recently announced about observation of two resonant
structures $X_{0}(2900)$ and $X_{1}(2900)$ (hereafter $X_{0}$ and $X_{1}$,
respectively) in the invariant $D^{-}K^{+}$ mass distribution of the decay $%
B^{+}\rightarrow D^{+}D^{-}K^{+}$ \cite{LHCb:2020}. In accordance with LHCb,
obtained results constitute the clear evidence for exotic mesons with full
open flavors. At the same time, the collaboration did not exclude models
which explain experimental data using hadronic rescattering effects.

The LHCb extracted the masses and widths of these structures, as well as
determined their spin-parities. Thus, it was shown that the $X_{0}$ is the
scalar resonance $J^{\mathrm{P}}=0^{+}$ with parameters%
\begin{eqnarray}
m_{0} &=&(2866\pm 7\pm 2)~\mathrm{MeV},\   \notag \\
\Gamma _{0} &=&(57\pm 12\pm 4)~\mathrm{MeV},  \label{eq:Data1}
\end{eqnarray}%
whereas $X_{1}$ is the vector state $J^{\mathrm{P}}=1^{-}$ and has the mass
and width
\begin{eqnarray}
m_{1} &=&(2904\pm 5\pm 1)~\mathrm{MeV},\   \notag \\
\Gamma _{1} &=&(110\pm 11\pm 4)~\mathrm{MeV}.  \label{eq:Data2}
\end{eqnarray}

From decay channels of these resonances $X_{0(1)}\rightarrow $ $D^{-}K^{+}$,
it is evident that they are built of four different valence quarks $ud%
\overline{s}\overline{c}$. These circumstances place $X_{0}$ and $X_{1}$ to
distinguishable position in the $XYZ$ family of exotic mesons. The LHCb's
discovery is doubly remarkable, because existence of the resonance $X(5568)$%
, seen by the D0 collaboration \cite{D0:2016mwd} and presumably composed of
quarks $sd\overline{b}\overline{u}$, was not later confirmed by other
experiments.

The LHCb information generated interesting theoretical investigations to
explain structure of new resonances $X_{0}$ and $X_{1}$, calculate their
masses, and if possible, estimate widths \cite%
{Karliner:2020vsi,Wang:2020xyc,Hu:2020mxp,Chen:2020aos,He:2020jna,Liu:2020orv, Liu:2020nil,Lu:2020qmp,Zhang:2020oze,Huang:2020ptc,Xue:2020vtq,Molina:2020hde}%
. In these papers the authors made different suggestions about internal
organization of these resonances, and used various methods and schemes to
compute their parameters. The diquark-antidiquark and hadronic molecule
pictures are dominant models to account for collected experimental data. For
example, in Refs.\ \cite{Karliner:2020vsi,Wang:2020xyc} $X_{0}$ was
considered as a scalar tetraquark $X_{0}=[sc][\overline{u}\overline{d}]$
using a phenomenological approach and the sum rule method, respectively. In
Ref.\ \cite{Chen:2020aos} $X_{0}$ was interpreted as $S$-wave hadronic $%
D^{\ast -}K^{\ast +}$ molecule, whereas $X_{1}$ considered $P$-wave
diquark-antidiquark state $X_{1}=[ud][\overline{s}\overline{c}]$. In Ref.\
\cite{Liu:2020orv}, on the contrary, it was asserted that two resonance-like
peaks generated in the process $B^{+}\rightarrow D^{+}D^{-}K^{+}$ due to
rescattering effects may emerge in LHCb experiment as the states\ $X_{0}$
and $X_{1}$.

It is worth noting that the exotic scalar meson with open flavor structure $%
X_{c}=[su][\overline{c}\overline{d}]$ was studied in our work \cite%
{Agaev:2016lkl}, in which it was explored as a charmed partner of the
resonance $X(5568)$. The mass and width of this tetraquark were calculated
using the sum rule method and two interpolating currents. These currents
correspond to structures $C\gamma _{5}\otimes \gamma _{5}C$ and $C\gamma
_{\mu }\otimes \gamma ^{\mu }C$, and are scalar-scalar (S) and axial-axial
currents (A), respectively. The width of $X_{c}$ was evaluated by analyzing
decay channels $X_{c}\rightarrow D_{s}^{-}\pi ^{+}$ and $X_{c}\rightarrow
D^{0}\overline{K}^{0}$. Performed studies led to the following results%
\begin{equation}
m_{\mathrm{S}}=(2634\pm 62)~\mathrm{MeV},\ \Gamma _{\mathrm{S}}=(57.7\pm
11.6)~\mathrm{MeV},  \label{eq:Pred1}
\end{equation}%
and
\begin{equation}
m_{\mathrm{A}}=(2590\pm 60)~\mathrm{MeV},\ \Gamma _{\mathrm{A}}=(63.4\pm
14.2)~\mathrm{MeV}.  \label{eq:Pred2}
\end{equation}%
The prediction $(2.55\pm 0.09)~\mathrm{GeV}$ for the mass of $X_{c}$ was
made in Ref.\ \cite{Chen:2016mqt}, as well.

It is clear that $X_{c}$ and $X_{0}$ are different particles and their decay
channels differ from each another. Nevertheless, it is convenient to compare
parameters of $X_{c}$ with LHCb data to make some preliminary assumptions on
structure of $X_{0}$. The mass of the ground-state tetraquark $X_{c}$ is not
large enough to account for LHCb data. We must also take into account that,
the tetraquark $X_{c}$ is composed of a relatively heavy diquark $[su]$ and
heavy antidiquark $[\overline{c}\overline{d}]$, whereas $X_{0}$ would has a
light diquark $[ud]$-heavy antidiquark $[\overline{c}\overline{s}]$
structure. Heavy-light tetraquarks are more compact and lighter than ones
with the same quark content but other compositions \cite{Agaev:2018khe}.
Therefore, the mass of the resonance $X_{0}$ should be within or even below
limits (\ref{eq:Pred1})-(\ref{eq:Pred2}) provided we treat it as a
ground-state tetraquark: In the diquark-antidiquark picture $X_{0}$ may be
considered as a radially excited $[ud][\overline{c}\overline{s}]$ state \cite%
{He:2020jna}.

Alternatively, one may analyze it as a hadronic molecule, i.e., as a bound
state of conventional $D$ and $K$ mesons. Mesons $D^{-}$ and $K^{+}$ may
form a bound state if the mass of a molecule $D^{-}K^{+}$ is less than
corresponding two-meson threshold $2365~\mathrm{MeV}$. But this estimate is
considerably below the mass of $X_{0}$, therefore, it is difficult to expect
that the molecule $D^{-}K^{+}$ can be considered as the $X_{0}$ state. For
compounds $\overline{D}^{\ast 0}K^{\ast 0}$ (hereafter $\overline{D}^{\ast
}K^{\ast }$) and $D^{\ast -}K^{\ast +}$ relevant two-particle thresholds are
equal to $\sim 2900~\mathrm{MeV}$, and hence they cannot dissociate to
vector mesons $D^{\ast }$ and $K^{\ast }$ provided masses of these molecules
are below this limit: An estimation for the mass $2848~\mathrm{MeV}$ of the
scalar molecule $D^{\ast }\overline{K}^{\ast }$ obtained in Ref.\ \cite%
{Molina:2010tx} supports this assumption. But such hadronic molecules can
decay to a pair of pseudoscalar $D$ and $K$ mesons. Then, structures $%
\overline{D}^{\ast }K^{\ast }$ and $D^{\ast -}K^{\ast +}$ may be interpreted
as $X_{0}$ if their masses are compatible with $m_{0}$. The scenario with $%
D^{\ast -}K^{\ast +}$ as $X_{0}$ was realized in Ref.\ \cite{Chen:2020aos},
in which the authors calculated the mass of the molecule $D^{\ast -}K^{\ast
+}$ in the framework of the QCD sum rule method. Result obtained there $%
2.87_{-0.14}^{+0.19}~\mathrm{GeV}$ indicates that an assumption about
molecule structure of $X_{0}$ deserves detailed studies.

In the present work, we treat $X_{0}$ as a hadronic molecule $\overline{D}%
^{\ast }K^{\ast }$ composed of two vector mesons $\overline{D}^{\ast 0}=%
\overline{c}u$ and $K^{\ast 0}=d\overline{s}$, and compute not only its
mass, but also width. The mass of $X_{0}$ is evaluated in the context of the
sum rule method, where we take into account quark, gluon and mixed
condensates up to dimension $15$. The width of $X_{0}$ is found by
considering the decay channel $X_{0}\rightarrow D^{-}K^{+}$. To this end, we
calculate the coupling $G$ that describes strong vertex $X_{0}D^{-}K^{+}$ in
the context of the light-cone sum rule (LCSR) approach using technical tools
of the soft-meson approximation. Information on this coupling obtained from
analysis allows us to estimate the width of $X_{0}$.

This work is structured in the following manner: In Section \ref{sec:Masses}%
, we calculate the mass and coupling of the hadronic molecule $\overline{D}%
^{\ast }K^{\ast }$. In Section \ref{sec:Decays}, we compute the strong
coupling $G$ by employing the LCSR method and soft-meson technique. In this
section we find also the width of the decay $X_{0}\rightarrow D^{-}K^{+}$.
Section \ref{sec:Disc} is reserved for discussion and our conclusions.


\section{Spectroscopic parameters of the $\overline{D}^{\ast }K^{\ast }$}

\label{sec:Masses}
The mass $m$, and current coupling $f$ of the hadronic molecule $\overline{D}%
^{*}K^{*}$ are necessary to check the assumption about a molecule nature of
the resonance $X_{0}$. These spectroscopic parameters are required also to
study its strong decay. We compute $m$, and $f$ using the QCD two-point sum
rule method \cite{Shifman:1978bx,Shifman:1978by}, which is one of the
effective nonperturbative approaches to determine parameters of the ordinary
and exotic hadrons.

The required sum rules can be derived from analysis of the two-point
correlation function
\begin{equation}
\Pi (p)=i\int d^{4}xe^{ipx}\langle 0|\mathcal{T}\{J(x)J^{\dag
}(0)\}|0\rangle,  \label{eq:CF1}
\end{equation}%
where $\mathcal{T}$ denotes the time-ordered product and $J(x)$ is the
interpolating current for the scalar particle~$X_{0}$. For molecule state $%
\overline{D}^{*}K^{*}$ this current is given by the expression
\begin{equation}
J(x)=[\overline{c}_{a}(x)\gamma ^{\mu }u_{a}(x)][\overline{s}_{b}(x)\gamma
_{\mu }d_{b}(x)].  \label{eq:CR1}
\end{equation}%
In Eq.\ (\ref{eq:CR1}) $c(x)$, $s(x)$, $u(x)$ and $d(x)$ stand for the
corresponding quark fields, whereas $a$ and $b$ are color indices.

Within the sum rule method the masses of various tetraquarks were analyzed
in numerous articles (see, for example, the review papers \cite%
{Chen:2016qju,Chen:2016spr,Albuquerque:2018jkn,Agaev:2020zad}), therefore
below we present only crucial points of performed analysis. In the sum rule
method the correlation function $\Pi (p)$ should be expressed both in terms
of physical parameters of $X_{0}$ and quark-gluon degrees of freedom. In the
first case, one finds the phenomenological side of the sum rules $\Pi ^{%
\mathrm{Phys}}(p)$ from Eq.\ (\ref{eq:CF1}) by inserting a complete set of
intermediate states. As a result we get
\begin{equation}
\Pi ^{\mathrm{Phys}}(p)=\frac{\langle 0|J|X_{0}\rangle \langle
X_{0}|J^{\dagger }|0\rangle }{m^{2}-p^{2}}+\cdots,  \label{eq:Phys1}
\end{equation}%
where the contribution of only the ground-state particle $X_{0}$ is shown
explicitly: Dots denote effects of higher resonances and continuum states in
the $X_{0}$ channel.

We have approximated the phenomenological side of the sum rule $\Pi ^{%
\mathrm{Phys}}(p)$ in Eq.\ (\ref{eq:Phys1}) using a simple-pole term. But,
in the case of the multiquark systems, this approach must be applied with
some caution, because the physical side may receive a contribution also from
two-hadron reducible terms. In fact, the relevant interpolating current
couples not only to the multiquark hadron, but interacts also with two
conventional hadron states lying below the mass of the multiquark system
\cite{Kondo:2004cr,Lee:2004xk}. These contributions can be either subtracted
from the sum rules or included into parameters of the pole term. In the case
of the tetraquarks the second approach is preferable and was applied in
articles \cite{Wang:2015nwa,Agaev:2018vag,Sundu:2018nxt}. It appears that,
the two-meson states generate the finite width $\Gamma (p)$ of the
tetraquark and lead to modification
\begin{equation}
\frac{1}{m^{2}-p^{2}}\rightarrow \frac{1}{m^{2}-p^{2}-i\sqrt{p^{2}}\Gamma (p)%
}.  \label{eq:Mod}
\end{equation}%
These effects, properly taken into account in the sum rules, rescale the
coupling $f$ \ leaving stable the mass $m$ of the tetraquark. Detailed
analyses proved that two-hadron contributions as a whole, and two-meson ones
in particular are small, and can be neglected \cite%
{Lee:2004xk,Wang:2015nwa,Agaev:2018vag,Sundu:2018nxt}. Therefore, to derive
the phenomenological side of the sum rules, we use in Eq.\ (\ref{eq:Phys1})
the zero-width single-pole approximation.

Introducing the spectroscopic parameters of $X_{0}$ through the matrix
element
\begin{equation}
\langle 0|J|X_{0}\rangle =fm,  \label{eq:ME1}
\end{equation}%
we rewrite $\Pi ^{\mathrm{Phys}}(p)$ in the final form%
\begin{equation}
\Pi ^{\mathrm{Phys}}(p)=\frac{f^{2}m^{2}}{m^{2}-p^{2}}+\cdots.
\label{eq:Phen2}
\end{equation}%
The function $\Pi ^{\mathrm{Phys}}(p)$ has a simple Lorentz structure
proportional to $\sim I$, and the term in Eq.\ (\ref{eq:Phen2}) is the
invariant amplitude $\Pi ^{\mathrm{Phys}}(p^{2})$ corresponding to this
structure.

The second component of the sum rules $\Pi ^{\mathrm{OPE}}(p)$, is
calculated in the operator product expansion ($\mathrm{OPE}$) with some
accuracy. To find $\Pi ^{\mathrm{OPE}}(p)$, we employ the explicit
expression of the interpolating current $J(x)$ in Eq.\ (\ref{eq:CF1}), and
contract relevant heavy and light quark fields. After these manipulations,
for $\Pi ^{\mathrm{OPE}}(p)$ we get
\begin{eqnarray}
&&\Pi ^{\mathrm{OPE}}(p)=i\int d^{4}xe^{ipx}\mathrm{Tr}\left[ \gamma ^{\mu
}S_{u}^{aa^{\prime }}(x)\gamma ^{\nu }S_{c}^{a^{\prime }a}(-x)\right]  \notag
\\
\times &&\mathrm{Tr}\left[ \gamma _{\mu }S_{d}^{bb^{\prime }}(x)\gamma _{\nu
}S_{s}^{b^{\prime }b}(-x)\right],  \label{eq:QCD1}
\end{eqnarray}%
where $S_{c}(x)$ and $S_{u(s,d)}(x)$ are the heavy $c$- and light $u(s,d)$%
-quark propagators, respectively. Their explicit expressions can be found,
for instance, in Ref.\ \cite{Agaev:2020zad}. The correlation function has
also a trivial structure: We denote the relevant invariant amplitude by $\Pi
^{\mathrm{OPE}}(p^{2})$.

The sum rules for $m$ and $f$ can be found by equating $\Pi ^{\mathrm{Phys}%
}(p^{2})$ and $\Pi ^{\mathrm{OPE}}(p^{2})$ and performing standard
manipulations of the method. First of all, one should apply the Borel
transformation to both sides of this equality to suppress contributions of
higher resonances and continuum states. At the next stage, by using the
hypothesis about quark-hadron duality, one subtracts higher resonances and
continuum terms from the physical side of the equality. As a result, the sum
rule equality starts to depend on the Borel $M^{2}$ and continuum threshold $%
s_{0}$ parameters.

The second equality required to find sum rules is obtained by applying the
operator $d/d(-1/M^{2})$ to the first expression. Then the sum rules for $m$
and $f$ are
\begin{equation}
m^{2}=\frac{\Pi ^{\prime }(M^{2},s_{0})}{\Pi (M^{2},s_{0})},  \label{eq:Mass}
\end{equation}%
and
\begin{equation}
f^{2}=\frac{e^{m^{2}/M^{2}}}{m^{2}}\Pi (M^{2},s_{0}).  \label{eq:Coupl}
\end{equation}%
Here, $\Pi (M^{2},s_{0})$ is the Borel transformed and subtracted invariant
amplitude $\Pi ^{\mathrm{OPE}}(p^{2})$, and $\Pi ^{\prime
}(M^{2},s_{0})=d/d(-1/M^{2})\Pi (M^{2},s_{0})$.

The function $\Pi (M^{2},s_{0})$ has the following form%
\begin{equation}
\Pi (M^{2},s_{0})=\int_{\mathcal{M}^{2}}^{s_{0}}ds\rho ^{\mathrm{OPE}%
}(s)e^{-s/M^{2}}+\Pi (M^{2}).  \label{eq:InvAmp}
\end{equation}%
Throughout this article, we neglect the mass of the quarks $u$ and $d$, and
set $\mathcal{M}=m_{c}+m_{s}$ in Eq.\ (\ref{eq:InvAmp}). The two-point
spectral density $\rho ^{\mathrm{OPE}}(s)$ is computed as an imaginary part
some of terms in the correlation function $\Pi ^{\mathrm{OPE}}(p)$. The
component $\Pi (M^{2})$ is the Borel transformation of remaining terms in $%
\Pi ^{\mathrm{OPE}}(p)$, and are obtained directly from their expressions.
Calculations are carried out by taking into account vacuum condensates up to
dimension $15$. The dimension-$15$ contribution to the correlation function
is proportional to product of light quark condensates $\langle \overline{s}%
g_{s}\sigma Gs\rangle \langle \overline{q}g_{s}\sigma Gq\rangle ^{2}$. This
and other higher dimensional terms in $\Pi (M^{2},s_{0})$ are obtained as
products of basic vacuum condensates using factorization procedure by
assuming that it does not lead to essential ambiguities. Analytical
expressions of $\rho ^{\mathrm{OPE}}(s)$ and $\Pi (M^{2})$ are rather
lengthy to be presented here explicitly.

The sum rules (\ref{eq:Mass}) and (\ref{eq:Coupl}) depend on universal quark
$\langle \overline{q}q\rangle =\langle 0|\overline{q}q|0\rangle $, gluon $%
\langle \alpha _{s}G^{2}/\pi \rangle =\langle 0|\alpha _{s}G^{2}/\pi
|0\rangle $ and mixed quark-gluon $\langle \overline{q}g_{s}\sigma Gq\rangle
=\langle 0|\overline{q}g_{s}\sigma Gq|0\rangle $ vacuum condensates ($%
q\equiv u,~d$, and similar expressions for the strange quark $s$) \cite%
{Shifman:1978bx,Shifman:1978by,Ioffe:1981kw,Ioffe:2005ym} and masses of $c$
and $s$ quarks%
\begin{eqnarray}
&&\langle \overline{q}q\rangle =-(0.24\pm 0.01)^{3}~\mathrm{GeV}^{3},\
\langle \overline{s}s\rangle =(0.8\pm 0.1)\langle \overline{q}q\rangle,
\notag \\
&&\langle \overline{q}g_{s}\sigma Gq\rangle =m_{0}^{2}\langle \overline{q}%
q\rangle,\ \langle \overline{s}g_{s}\sigma Gs\rangle =m_{0}^{2}\langle
\overline{s}s\rangle,  \notag \\
&&m_{0}^{2}=(0.8\pm 0.2)~\mathrm{GeV}^{2}  \notag \\
&&\langle \frac{\alpha _{s}G^{2}}{\pi }\rangle =(0.012\pm 0.004)~\mathrm{GeV}%
^{4},  \notag \\
&&m_{s}=93_{-5}^{+11}~\mathrm{MeV},\ m_{c}=1.27\pm 0.2~\mathrm{GeV}.
\label{eq:Parameters}
\end{eqnarray}%
As is seen, the vacuum condensate of strange quarks is different from $%
\langle 0|\overline{q}q|0\rangle $ \cite{Ioffe:1981kw}. The mixed
condensates $\langle \overline{q}g_{s}\sigma Gq\rangle $ and $\langle
\overline{s}g_{s}\sigma Gs\rangle $ can be expressed in terms of the
corresponding quark condensates and parameter $m_{0}^{2}$, numerical value
of which was extracted from analysis of baryonic resonances \cite%
{Ioffe:2005ym}.

\begin{widetext}

\begin{figure}[h!]
\begin{center}
\includegraphics[totalheight=6cm,width=8cm]{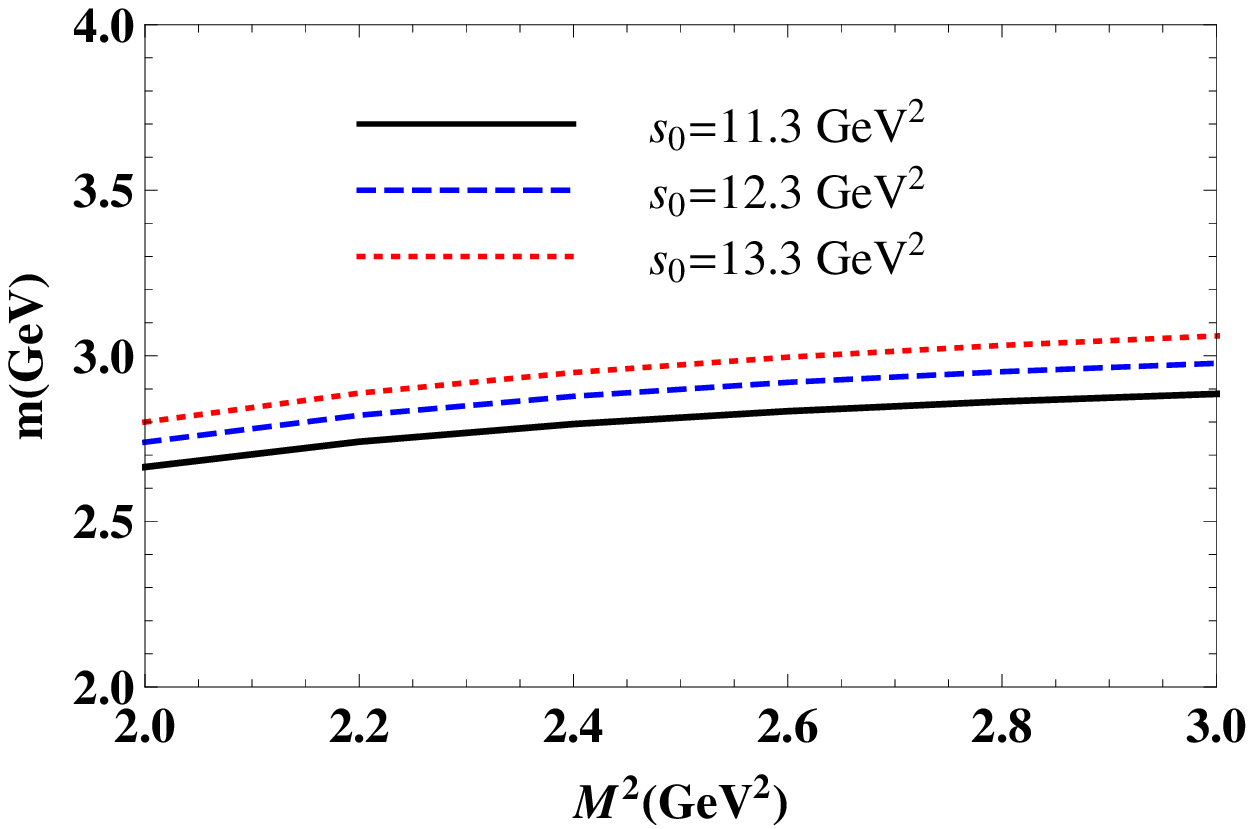}
\includegraphics[totalheight=6cm,width=8cm]{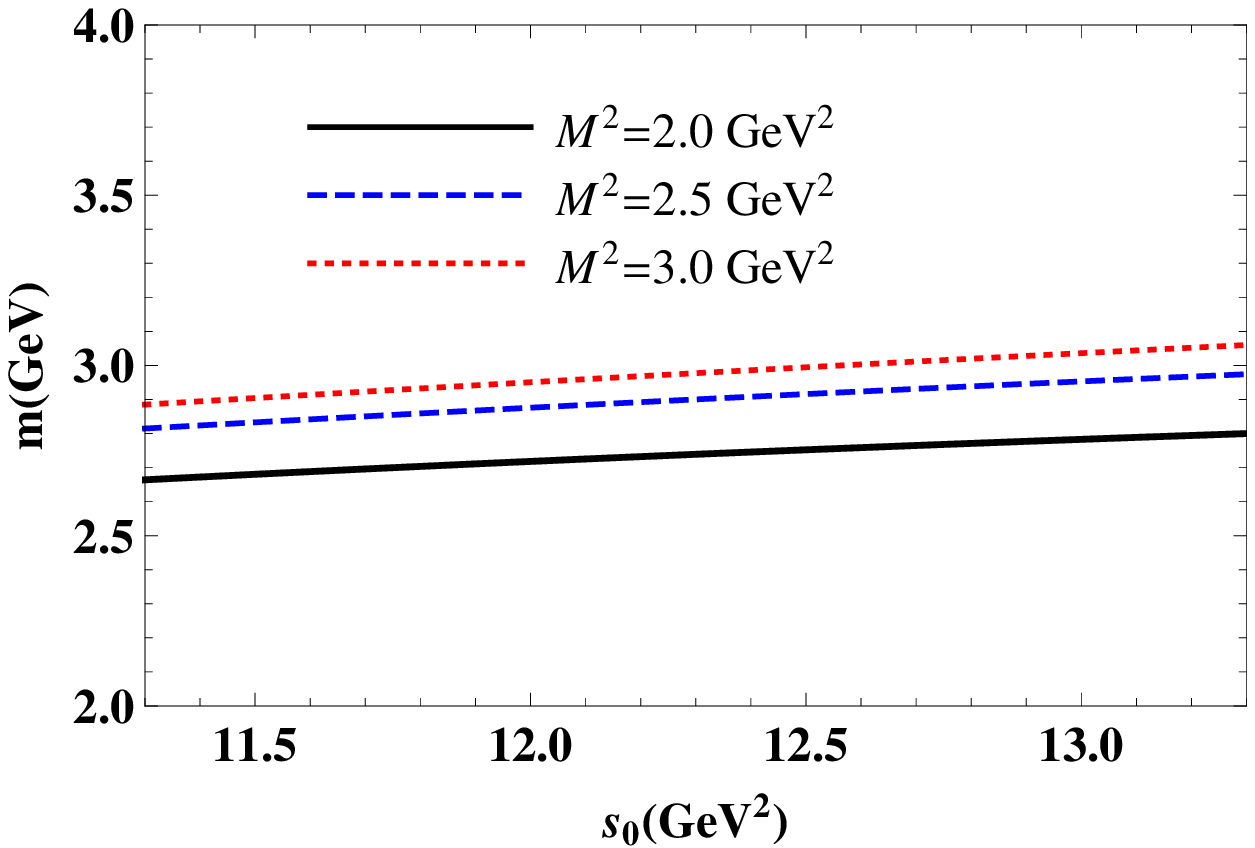}
\end{center}
\caption{The mass $m$ of the $X_0$ as a function of the Borel parameter $M^{2}$ at fixed $s_{0}$ (left panel), and as a function of the continuum threshold parameter $s_0$ at fixed $M^2$ (right panel).}
\label{fig:Mass}
\end{figure}

\end{widetext}

The $m$ and $f$ are functions of the parameters $M^{2}$ and $s_{0}$, as
well. The correct choice for $M^{2}$ and $s_{0}$ is an important problem of
sum rule computations. The working regions for $M^{2}$ and $s_{0}$ should
satisfy usual constraints imposed on the pole contribution ($\mathrm{PC}$)
and convergence of the operator product expansion. To analyze these
questions, we introduce the quantities
\begin{equation}
\mathrm{PC}=\frac{\Pi (M^{2},s_{0})}{\Pi (M^{2},\infty )},  \label{eq:PC}
\end{equation}%
and
\begin{equation}
R(M^{2})=\frac{\Pi ^{\mathrm{DimN}}(M^{2},s_{0})}{\Pi (M^{2},s_{0})}.
\label{eq:Convergence}
\end{equation}%
In Eq.\ (\ref{eq:Convergence}) $\Pi ^{\mathrm{DimN}}(M^{2},s_{0})$ is a last
term (or a sum of last few terms) in the correlation function. In the
present analysis, we use the sum of last three terms in $\mathrm{OPE}$, and
hence $\mathrm{DimN\equiv Dim(13+14+15)}$.

The $\mathrm{PC}$ is used to fix upper bound for $M^{2}$, whereas $R(M^{2})$
is necessary to find low limit for the Borel parameter. These two values of $%
M^{2}$ fix the boundaries of a region where the Borel parameter can be
varied. Our analysis shows that the working regions for the parameters $%
M^{2} $ and $s_{0}$ are
\begin{equation}
M^{2}\in \lbrack 2,3]\ \mathrm{GeV}^{2},\ s_{0}\in \lbrack 11.3,13.3]\
\mathrm{GeV}^{2},  \label{eq:Wind1}
\end{equation}%
and they obey restrictions on $\mathrm{PC}$ and convergence of $\mathrm{OPE}$%
. Thus, at $M^{2}=3~\mathrm{GeV}^{2}$ the pole contribution is $0.5$,
whereas at $M^{2}=2~\mathrm{GeV}^{2}$ it becomes equal to $0.8$. At the
minimum of $M^{2}=2~\mathrm{GeV}^{2}$, we find $R\approx 0.01$, which
guarantees the convergence of the sum rules. We extract the parameters $m$
and $f$ approximately at a middle point of the window (\ref{eq:Wind1}), $%
M^{2} =2.5~\mathrm{GeV}^{2}$ and $s_0=12~\mathrm{GeV}^{2}$, where the pole
contribution is $\mathrm{PC}\approx 0.65$. This fact ensures the ground
state nature of $X_{0}$.

Our predictions for $m$ and $f$ are
\begin{eqnarray}
m &=&(2868~\pm 198)~\mathrm{MeV},  \notag \\
f &=&(3.0\pm 0.7)\times 10^{-3}~\mathrm{GeV}^{4}.  \label{eq:Result1}
\end{eqnarray}

The sum rule results, in general, should not depend on the parameter $M^{2}$. 
But in real calculations $m$ and $f$ are sensitive to the choice of $M^{2}$. 
From inspection of Eq.\ (\ref{eq:Result1}) it is seen, that theoretical
uncertainties in the case of $m$ equal to $\pm 6.9\%$, whereas for the
coupling $f$ they amount to $\pm 23.3\%$. Theoretical ambiguities of $f$ are
larger, because $f$ is determined directly in terms of $\Pi (M^{2},s_{0})$,
whereas the sum rule for $m$ depends on the ratio of such functions and is
exposed to smaller variations. Nevertheless, uncertainties even for the
coupling $f$ remain within limits accepted in sum rule computations. In
Fig.\ \ref{fig:Mass} we display the sum rule's predictions for $m$ as a
function of $M^{2}$, where one can see its residual dependence on the Borel
parameter.

The continuum threshold parameter $s_{0}$ separates a ground-state
contribution from effects due to higher resonances and continuum states. In
other words, $\sqrt{s_{0}}$ has to be smaller than the mass $m^{\ast }$ of
the first excitation of the $X_{0}$. The self-consistent sum rule analysis
implies that the difference $\sqrt{s_{0}}-m$ is around or less than $m^{\ast
}-m$. Excited states of conventional hadrons and their parameters are known
either from experimental measurements or from alternative theoretical
studies. In the case of the multiquark hadrons there is a deficiency of
relevant information. The mass gap $\sqrt{s_{0}}-m\approx (500-600)~\mathrm{%
MeV}$ found in the present work can be considered as a reasonable estimate $%
m^{\ast }\approx (m+500)~\mathrm{MeV}$ for the hadronic molecule $\overline{D%
}^{\ast }K^{\ast }$ containing one heavy quark. Dependence of extracted
value of $m$ on the scale $s_{0}$ is also shown in Fig.\ \ref{fig:Mass}.

Obtained prediction for the mass of the state $\overline{D}^{\ast }K^{\ast }$
is in a nice agreement with new LHCb measurements. This is necessary, but
not enough to interpret $X_{0}$ as the hadronic molecule. For reliable
conclusions, we need to calculate the width of the molecule $\overline{D}%
^{\ast }K^{\ast }$ and confront it with data: only together these parameters
can support or not assumptions about the structure of $X_{0}$.


\section{The decay $X_{0}\rightarrow D^{-}K^{+}$}

\label{sec:Decays}

In this section we explore the strong decay $X_{0}\rightarrow D^{-}K^{+}$ in
order to find width of the resonance $X_{0}$. Strictly speaking, there are
other decay channel of the scalar state $X_{0}$, namely the $S$-wave decay
to a pair of mesons $\overline{D}^{0}K^{0}$. Because $X_{0}$ was observed as
enhancement in the $D^{-}K^{+}$ mass distribution, we concentrate on the
first process and saturate full width of the resonance $X_{0}$ by this
channel.

The width of the decay $X_{0}\rightarrow D^{-}K^{+}$ is determined by the
strong coupling $G$ corresponding to the vertex $X_{0}D^{-}K^{+}$. We are
going to calculate $G$ using method of the QCD sum rule on the light-cone
\cite{Balitsky:1989ry,Belyaev:1994zk} and methods of the soft-meson
approximation \cite{Ioffe:1983ju}. To this end, we start from analysis of
the correlation function \cite{Belyaev:1994zk,Cragie:1982ng}
\begin{equation}
\Pi (p,q)=i\int d^{4}xe^{ipx}\langle K(q)|\mathcal{T}\{J^{D}(x)J^{\dag
}(0)\}|0\rangle,  \label{eq:CorrF3}
\end{equation}%
where by $K$ and $D$ we denote, in short forms, the mesons $K^{+}$ and $%
D^{-} $, respectively. In the correlation function $\Pi (p,q)$, the
interpolating current $J(x)$ is given by Eq.\ (\ref{eq:CR1}), whereas for $%
J^{D}(x)$ we use
\begin{equation}
J^{D}(x)=\overline{c}_{n}(x)i\gamma _{5}d_{n}(x),  \label{eq:Dcur}
\end{equation}%
with $n$ being the color index. It is not difficult to determine $\Pi (p,q)$
in terms of the physical parameters of the particles involved into the decay
\cite{Belyaev:1994zk}. By taking into account the ground states in the $D$
and $X_{0}$ channels, we get
\begin{eqnarray}
\Pi ^{\mathrm{Phys}}(p,q) &=&\frac{\langle 0|J^{D}|D\left( p\right) \rangle
}{p^{2}-m_{D}^{2}}\langle D\left( p\right) K(q)|X_{0}(p^{\prime })\rangle
\notag \\
&&\times \frac{\langle X_{0}(p^{\prime })|J^{\dagger }|0\rangle }{p^{\prime
2}-m^{2}}+\cdots,  \label{eq:CorrF4}
\end{eqnarray}%
where $p$, $q$ and $p^{\prime }=p+q$ are the momenta of the particles $D$, $%
K $, and $X_{0}$, and $m_{D}$ is the mass of $D$ meson. The ellipses in Eq.\
(\ref{eq:CorrF4}) refer to contributions of higher resonances and continuum
states in the $D$ and $X_{0}$ channels.

In order to finish computation of the correlation function, we introduce the
matrix elements
\begin{eqnarray}
&&\langle 0|J^{D}|D\left( p\right) \rangle =\frac{f_{D}m_{D}^{2}}{m_{c}},\
\langle X_{0}(p^{\prime })|J^{\dagger }|0\rangle =fm,  \notag \\
&&\langle D\left( p\right) K(q)|X_{0}(p^{\prime })\rangle =Gp\cdot p^{\prime
}.  \label{eq:Mel}
\end{eqnarray}%
In expressions above $f_{D}\,$ is decay constant of the meson $D^{-}$. Then
for $\Pi ^{\mathrm{Phys}}(p,q)$ we find%
\begin{equation}
\Pi ^{\mathrm{Phys}}(p,q)=\frac{Gfmf_{D}m_{D}^{2}}{%
m_{c}(p^{2}-m_{D}^{2})(p^{\prime 2}-m^{2})}p\cdot p^{\prime }+\cdots.
\label{eq:CorrF5}
\end{equation}

To continue, we have to calculate $\Pi ^{\mathrm{QCD}}(p,q)$ in terms of the
quark-gluon degrees of freedom and find the QCD side of the sum rule.
Contractions in Eq.\ (\ref{eq:CorrF3}) of $c$ and $d$ quark and antiquark
fields yield
\begin{eqnarray}
&&\Pi ^{\mathrm{OPE}}(p,q)=\int d^{4}xe^{ipx}\left[ \gamma ^{\mu
}S_{c}^{bn}(-x){}\gamma _{5}\right.  \notag \\
&&\left. \times S_{d}^{na}(x){}\gamma _{\mu }\right] _{\alpha \beta }\langle
K(q)|\overline{u}_{\alpha }^{b}(0)s_{\beta }^{a}(0)|0\rangle,
\label{eq:CorrF6}
\end{eqnarray}%
where $\alpha $ and $\beta $ are the spinor indexes.

Apart from quark propagators the function $\Pi ^{\mathrm{OPE}}(p,q)$ depends
also on local matrix elements of the quark operator $\overline{u}s$
sandwiched between the vacuum and $K$ meson. To express $\langle K(q)|%
\overline{u}_{\alpha }^{b}(0)s_{\beta }^{a}(0)|0\rangle $ using the $K$
meson's local matrix elements, we expand $\overline{u}(0)s(0)$ over the full
set of Dirac matrices $\Gamma ^{J}$ and project them onto the color-singlet
states
\begin{equation}
\overline{u}_{\alpha }^{b}(0)s_{\beta }^{a}(0)\rightarrow \frac{1}{12}\delta
^{ba}\Gamma _{\beta \alpha }^{J}\left[ \overline{u}(0)\Gamma ^{J}s(0)\right]
,  \label{eq:MatEx}
\end{equation}%
where
\begin{equation}
\Gamma ^{J}=\mathbf{1},\ \gamma _{5},\ \gamma _{\mu },\ i\gamma _{5}\gamma
_{\mu },\ \sigma _{\mu \nu }/\sqrt{2}.  \label{eq:Dirac}
\end{equation}

The expression (\ref{eq:MatEx}) reveals a main difference between vertices
composed of conventional mesons and vertices containing a tetraquark and two
ordinary mesons. Indeed, in the vertices of ordinary mesons the correlation
function depends on distribution amplitudes (DAs) of one of the final-state
mesons, for example, on DAs of $K$ meson. The DAs of the mesons are
determined as non-local matrix elements of relevant quark fields placed
between the meson and vacuum states. In the case under discussion, it is
evident that instead of non-local matrix elements, we have $\Pi ^{\mathrm{QCD%
}}(p,q)$ that contains local matrix elements of $K$ meson. The reason is
that $X_{0}$ and interpolating current Eq.\ (\ref{eq:CR1}) are built of four
quark fields at the same space-time location. Substitution of this current
into the correlation function and contractions of $c$ and $d$ quark fields
yield expressions, where the remaining light quarks are sandwiched between
the $K$ meson and vacuum states, forming local matrix elements. In other
words, we encounter the situation when dependence of the correlation
function on the meson DAs disappears and integrals over the meson DAs reduce
to overall normalization factors. In the framework of the LCSR method such
situation is possible in the kinematical limit $q\rightarrow 0$, when the
light-cone expansion is replaced by the short-distant one. As a result,
instead of the expansion in terms of DAs, one gets expansion over the local
matrix elements \cite{Belyaev:1994zk}. The limit $q\rightarrow 0$ is known
as the soft-meson approximation. In this approximation $p=p^{\prime }$ and
invariant amplitudes $\Pi ^{\mathrm{Phys}}(p^{2})$ and $\Pi ^{\mathrm{OPE}%
}(p^{2})$ depend only on the variable $p^{2}$. For our purposes a decisive
fact is the observation made in Ref.\ \cite{Belyaev:1994zk}: the soft-meson
approximation and full LCSR treatment of the conventional mesons' vertices
lead to predictions which are numerically very close to each other.

The soft-meson approximation simplifies the QCD component of the sum rule,
but leads to additional complications in its phenomenological side. In this
limit for invariant amplitude $\Pi ^{\mathrm{Phys}}(p^{2})$ we get%
\begin{equation}
\Pi ^{\mathrm{Phys}}(p^{2})=G\frac{fmf_{D}m_{D}^{2}\widetilde{m}^{2}}{%
m_{c}(p^{2}-\widetilde{m}^{2})^{2}}+\cdots,  \label{eq:CF2}
\end{equation}%
where $\widetilde{m}^{2}=(m^{2}+m_{D}^{2})/2$. This amplitude contains the
double pole at $p^{2}=\widetilde{m}^{2}$, therefore its Borel transformation
is given by the formula

\begin{eqnarray}
&&\Pi ^{\mathrm{Phys}}(M^{2})=G\frac{fmf_{D}m_{D}^{2}}{m_{c}}\frac{%
\widetilde{m}^{2}e^{-\widetilde{m}^{2}/M^{2}}}{M^{2}}+\cdots.  \notag \\
&&  \label{eq:CF3}
\end{eqnarray}

In the standard approach the invariant amplitude $\Pi ^{\mathrm{Phys}%
}(p^{2},p^{\prime 2})$ depends on two variables $p^{2}$ and $p^{\prime 2}$,
and the Borel transformations over $p^{2}$ and $p^{\prime 2}$ suppress
contributions of higher resonances and continuum states. The suppressed
terms afterwards can be subtracted using assumption on quark-hadron duality.
But in the soft limit even after Borel transformation besides ground-state
term $\Pi ^{\mathrm{Phys}}(M^{2})$ contains additional unsuppressed
contributions. This is a price to be paid in the soft approximation for
simple $\Pi ^{\mathrm{OPE}}(p^{2})$ expression. To remove contaminating
contributions from the phenomenological side of the sum rule, one has to act
on $\Pi ^{\mathrm{Phys}}(M^{2})$ by the operator \cite%
{Belyaev:1994zk,Ioffe:1983ju}
\begin{equation}
\mathcal{P}(M^{2},m^{2})=\left( 1-M^{2}\frac{d}{dM^{2}}\right)
M^{2}e^{m^{2}/M^{2}},  \label{eq:Oper}
\end{equation}%
that singles out the ground-state term. Contributions remaining in $\Pi ^{%
\mathrm{Phys}}(M^{2})$ after this prescription can be subtracted by the
usual way. Then the sum rule for the strong coupling $G$ reads%
\begin{equation}
G=\frac{m_{c}\widetilde{m}^{2}}{fmf_{D}m_{D}^{2}}\mathcal{P}(M^{2},%
\widetilde{m}^{2})\Pi ^{\mathrm{OPE}}(M^{2},s_{0}).  \label{eq:SRcoupl}
\end{equation}

Returning to the calculation of $\Pi ^{\mathrm{OPE}}(p,q)$, it is worth
noting that by substituting the expansion (\ref{eq:MatEx}) into Eq.\ (\ref%
{eq:CorrF3}), one has to perform summations over color indices and fix local
matrix elements of $K$ meson that contribute to $\Pi ^{\mathrm{OPE}}(p^{2})$
in the soft limit. There are only a few such elements: Two-particle matrix
elements of twist-2 and twist-3
\begin{eqnarray}
&&\langle 0|\overline{u}\gamma _{\mu }\gamma _{5}s|K(q)\rangle =if_{K}q_{\mu
},  \notag \\
&&\langle 0|\overline{u}i\gamma _{5}s|K\rangle =\frac{f_{K}m_{K}^{2}}{m_{s}}.
\label{eq:MatElK1}
\end{eqnarray}%
There are also three-particle local matrix elements of $K$ meson, for an
example,%
\begin{equation}
\langle 0|\overline{u}\gamma ^{\nu }\gamma _{5}igG_{\mu \nu }s|K(q)\rangle
=iq_{\mu }f_{K}m_{K}^{2}\kappa _{4K},  \label{eq:MatElK2}
\end{equation}%
which may contribute to $\Pi ^{\mathrm{OPE}}(p^{2})$. The elements given by
Eq.\ (\ref{eq:MatElK1}) appear due to the propagator $S_{d}$, perturbative
term of $S_{c}$, and the expansion (\ref{eq:MatEx}), whereas three-particle
matrix elements may contribute after gluon insertions to $\overline{u}\Gamma
^{J}s$ stemming from nonperturbative components of the propagator $S_{c}$.
In matrix elements (\ref{eq:MatElK1}) and (\ref{eq:MatElK2}) quark and gluon
fields are fixed at the same position $x=0$, and $\kappa _{4K}$ is the
twist-4 matrix element of the $K$ meson.

Procedures to calculate the correlation function in the soft approximation
were presented in Refs.\ \cite{Agaev:2016ijz,Agaev:2016dev}, therefore we
skip further technical details and provide final expression for the
correlation function, which is calculated with dimension-$9$ accuracy and
given as a sum of the perturbative and nonperturbative components\
\begin{eqnarray}
\Pi ^{\mathrm{OPE}}(M^{2},s_{0}) &=&\frac{\mu _{K}}{8\pi ^{2}}\int_{\mathcal{%
M}^{2}}^{s_{0}}\frac{ds(m_{c}^{2}-s)^{2}}{s}e^{-s/M^{2}}+\Pi _{\mathrm{NP}%
}(M^{2}).  \notag \\
&&  \label{eq:DecayCF}
\end{eqnarray}%
The nonperturbative component of the correlation function $\Pi _{\mathrm{NP}%
}(M^{2})$ has the following form%
\begin{eqnarray}
&&\Pi _{\mathrm{NP}}(M^{2})=-\frac{\langle \overline{d}d\rangle \mu _{K}m_{c}%
}{3}e^{-m_{c}^{2}/M^{2}}  \notag \\
&&+\langle \frac{\alpha _{s}G^{2}}{\pi }\rangle \frac{\mu _{K}m_{c}^{4}}{%
72M^{4}}\int_{0}^{1}\frac{dz}{z^{3}(z-1)^{3}}e^{-m_{c}^{2}/[M^{2}z(1-z)]}
\notag \\
&&+\frac{\langle \overline{d}g\sigma Gd\rangle \mu _{K}m_{c}^{3}}{6M^{4}}%
e^{-m_{c}^{2}/M^{2}}-\langle \frac{\alpha _{s}G^{2}}{\pi }\rangle \langle
\overline{d}d\rangle  \notag \\
&&\times \frac{\mu _{K}m_{c}(m_{c}^{2}+3M^{2})\pi ^{2}}{54M^{6}}%
e^{-m_{c}^{2}/M^{2}}+\langle \frac{\alpha _{s}G^{2}}{\pi }\rangle \langle
\overline{d}g\sigma Gd\rangle  \notag \\
&&\times \frac{\mu _{K}m_{c}(m_{c}^{4}+6M^{2}m_{c}^{2}+6M^{4})\pi ^{2}}{%
216M^{10}}e^{-m_{c}^{2}/M^{2}}.  \label{eq:DecayNPCF}
\end{eqnarray}%
In expressions above, we introduce $\mu _{K}=f_{K}m_{K}^{2}/m_{s}$. It turns
out that only the twist-3 matrix element from Eq.\ (\ref{eq:MatElK1})
contributes to the function $\Pi _{\mathrm{NP}}(M^{2})$. The last term in $%
\Pi _{\mathrm{NP}}(M^{2})$ is proportional to $\langle \alpha _{s}G^{2}/\pi
\rangle \langle \overline{d}g\sigma Gd\rangle $ with dimension $9$, and is
suppressed additionally by the factor $1/M^{6}$. Therefore, dimension-$9$
accuracy for computation of $\Pi ^{\mathrm{OPE}}(M^{2},s_{0})$ adopted in
the present article is high and enough to get reliable result.

\begin{table}[tbp]
\begin{tabular}{|c|c|}
\hline\hline
Quantities & Values (in $\mathrm{MeV}$ units) \\ \hline
$m_{D}$ & $1869.61\pm 0.10$ \\
$m_{K}$ & $493.68\pm 0.02$ \\
$f_{D}$ & $211.9 \pm 1.1$ \\
$f_{K}$ & $155.6 \pm 0.4$ \\ \hline\hline
\end{tabular}%
\caption{Parameters of the mesons $D^-$ and $K^+$ used in numerical
computations.}
\label{tab:Param}
\end{table}

The sum rule (\ref{eq:SRcoupl}) depends on the various vacuum condensates
written down above (\ref{eq:Parameters}). It contains the masses and decay
constants of the final-state mesons $D^-$ and $K^+$: relevant spectroscopic
parameters are collected in Table\ \ref{tab:Param}. All of them are borrowed
from Ref.\ \cite{Tanabashi:2018oca}. For decay constants $f_D$ and $f_K$
Particle Data Group's averages are used.

To carry out numerical analysis one also needs to fix $M^{2}$ and $s_{0}$.
The restrictions imposed on these auxiliary parameters are standard for sum
rule computations and have been discussed above. Our analysis demonstrates
that working regions (\ref{eq:Wind1}) meet all constraints necessary for
computations of $\Pi ^{\mathrm{OPE}}(M^{2},s_{0})$. Numerical calculations
lead to the following result
\begin{equation}
|G|=(0.66\pm 0.06)\ \mathrm{GeV}^{-1}.  \label{eq:Coupl1}
\end{equation}%
This prediction for the strong coupling $G$ is typical for tetraquark-meson
vertices. Its numerical value and dimension are determined by definition of
matrix element $\langle D\left( p\right) K(q)|X_{0}(p^{\prime })\rangle $:\
Modification of the vertex $DKX_{0}$ in Eq.\ (\ref{eq:Mel}) changes the
value and dimension of $G$. In general, it is possible to rewrite $\langle
D\left( p\right) K(q)|X_{0}(p^{\prime })\rangle $ in such a way that to make
$G$ dimensionless. In our analysis $G$ is an intermediate parameter, whereas
the physical quantity to be found is the width $\Gamma $ of the decay $%
X_{0}\rightarrow D^{-}K^{+\text{ }}$. The $\Gamma $ is calculated by taking
into account Eq.\ (\ref{eq:Mel}), and its expression depends on these matrix
elements. But regardless used convention for the vertex $DKX_{0}$ and
analytical form of the width, numerical computations lead to the same final
result with correct dimension, as it should be for a physical quantity.

Having used the matrix elements given by Eq.\ (\ref{eq:Mel}), we derive for
the width of the decay $X_{0}\rightarrow D^{-}K^{+\text{ }}$
\begin{equation}
\Gamma \left[ X_{0}\rightarrow D^{-}K^{+\text{ }}\right] =\frac{%
G^{2}m_{D}^{2}\lambda }{8\pi }\left( 1+\frac{\lambda ^{2}}{m_{D}^{2}}\right)
,  \label{eq:DW}
\end{equation}%
where $\lambda =\lambda \left( m,m_{D},m_{K}\right) $ and
\begin{eqnarray}
\lambda \left( a,b,c\right) &=&\frac{1}{2a}\left( a^{4}+b^{4}+c^{4}\right.
\notag \\
&&\left. -2\left( a^{2}b^{2}+a^{2}c^{2}+b^{2}c^{2}\right) \right) ^{1/2}.
\end{eqnarray}%
Then it is not difficult to find that
\begin{equation}
\Gamma \left[ X_{0}\rightarrow D^{-}K^{+}\right] =(49.6\pm 9.3)~\mathrm{MeV}.
\label{eq:DW1Numeric}
\end{equation}%
This prediction for the width of the resonance $X_{0}$ is in a reasonable
agreement with new LHCb data (\ref{eq:Data1}).

The molecule $\overline{D}^{\ast 0}K^{\ast 0}$ is composed of two neutral
vector mesons, which are strong-interaction unstable particles. The meson $%
\overline{D}^{\ast }(2007)^{0}$ is relatively narrow state $\Gamma _{%
\overline{D}^{\ast }}<2.1~\mathrm{MeV}$, whereas the width $\Gamma _{K^{\ast
}}=(47.4\pm 0.5)~\mathrm{MeV}$ of $K^{\ast }(892)^{0}$ within experimental
errors is comparable with LHCb data $\Gamma _{0}$. In Ref.\ \cite%
{Chen:2020aos} $X_{0}$ was modeled as hadronic $D^{\ast -}K^{\ast +}$
molecule, and width of the meson $K^{\ast +}$ was used to estimate roughly
the $X_{0}$ resonance's width. The hadronic molecules $D^{\ast -}K^{\ast +}$
and $\overline{D}^{\ast 0}K^{\ast 0}$ are bound states, and partial widths
of their decay channels are determined by quark-gluon interactions inside of
these particles. Due to multiquark nature of molecules their internal
dynamics obviously differs from those of free mesons $D^{\ast }$ and $%
K^{\ast }$. Therefore, estimation of the $D^{\ast -}K^{\ast +}$ and $%
\overline{D}^{\ast 0}K^{\ast 0}$ molecules' widths using directly widths of
constituent mesons seems us to be problematic. One can suggest, that after
dissociation of $\overline{D}^{\ast 0}K^{\ast 0}$ to two vector mesons,
decays of $K^{\ast 0}$ may be used for such analysis. But the mass of the $%
X_{0}$ is below relevant two-meson thresholds in both pictures, i.e., $X_{0}$
does not decay to mesons $\overline{D}^{\ast 0}+K^{\ast 0}$ or $D^{\ast
-}+K^{\ast +}$.  The $P$-wave decay $\overline{D}^{\ast 0}K^{\ast
0}\rightarrow \overline{D}_{0}^{\ast }(2400)^{0}+K^{\ast 0}$ with the vector
meson $K^{\ast 0}$ in the final state is forbidden kinematically. Another
two-body $P$-wave decays of $\overline{D}^{\ast 0}K^{\ast 0}$, for example,
to mesons $\overline{D}^{\ast }(2007)^{0}+K_{0}^{\ast }(1430)$ and $%
\overline{D}_{1}(2420)^{0}+K^{0}$ are not allowed because of the same
reasons. \ Nevertheless, there are multibody decays of $X_{0}$ which
contribute to its full width. For example, processes $X_{0}\rightarrow
D^{-}K^{+}\pi ^{0}\pi ^{0}$ and $X_{0}\rightarrow D^{-}K^{+}\pi ^{+}\pi ^{-}$
can improve  our prediction for $\Gamma $. But these processes imply
production of $4$ new valence quarks through different mechanisms, which
suppress relevant amplitudes by the factor $\alpha _{s}$ or additional
strong couplings. As a result, widths of such subdominant decays would be
small.

We have explored the dominant decay channel $X_{0}\rightarrow D^{-}K^{+}$ of
the resonance $X_{0}$. The result for its width in Eq.\ (\ref{eq:DW1Numeric}%
) has been obtained in the context of the LCSR method by applying first
principles of the QCD, and is reliable prediction for this parameter. It can
be improved further by including into analysis other decay channels of the $%
X_{0}$, which are beyond the scope of the present article.


\section{Discussion and conclusions}

\label{sec:Disc}
In the present work we have explored one of two new resonances $X_{0}$ and $%
X_{1}$ reported by the LHCb collaboration. Namely, we have considered $X_{0}$
as a scalar hadronic molecule $\overline{D}^{\ast }K^{\ast }$, and
calculated its mass and width. For these purposes, we have used the QCD sum
rule method. The spectroscopic parameters of the state $\overline{D}^{\ast
}K^{\ast }$ have been extracted from two-point sum rules, whereas for
analysis of its strong decay channel, we used LCSR method and soft-meson
approximation. Obtained predictions for $m$ and $\Gamma $ are in nice
agreement with reported LHCb data, which can be interpreted in favor of
molecule nature of the resonance $X_{0}$.

The suggestion about molecular structure of $X_{0}$ was made in Ref.\ \cite%
{Chen:2020aos}, in which the authors computed the mass of the molecule $%
D^{\ast -}K^{\ast +}$ using the sum rule method. Calculations were done by
taking into account nonperturbative terms up to dimension $8$. Prediction
obtained there for $\widetilde{m}$
\begin{equation}
\widetilde{m}=2870_{-140}^{+190}~\mathrm{MeV}
\end{equation}%
is very close to our result. The molecule composition for the $X_{0}$ in
different frameworks was proposed in Refs.\ \cite%
{Hu:2020mxp,Liu:2020nil,Huang:2020ptc}, as well.

Alternatively, the resonance $X_{0}$ was analyzed in Refs. \cite%
{Karliner:2020vsi,Wang:2020xyc,He:2020jna,Zhang:2020oze} by assuming that it
is a diquark-antidiquark state. The sum rule prediction for the mass of the
scalar tetraquark $T_{1}=[sc][\overline{u}\overline{d}]$ with an axial-axial
$C\gamma _{\mu }\otimes \gamma ^{\mu }C$ type \ structure reads \cite%
{Wang:2020xyc}%
\begin{equation}
m_{T_{1}}=(2910\pm 120)~\mathrm{MeV}.
\end{equation}%
The similar sum rule investigations were performed in Ref.\ \cite%
{Zhang:2020oze}. Results for masses of the scalar tetraquark $T_{2}=[ud][%
\overline{s}\overline{c}]$ with scalar-scalar and axial-axial structures are
equal to%
\begin{eqnarray}
m_{T_{2S}} &=&2750_{-190}^{+180}~\mathrm{MeV},  \notag \\
m_{T_{2A}} &=&2770_{-200}^{+190}~\mathrm{MeV},
\end{eqnarray}%
respectively. By taking into account uncertainties of calculations, the
author concluded that $X_{0}$ could be interpreted as a tetraquark $J^{%
\mathrm{P}}=0^{+}$ with either scalar-scalar or axial-axial configurations.
It is seen that states $T_{1}$ and $T_{2}$ are connected by the relations $%
T_{1}=\overline{T_{2}}$ or $\overline{T_{1}}=T_{2}$ as conventional mesons,
for example, $D^{0}=c\overline{u}$ and $\overline{D^{0}}=\overline{c}u$.
Masses of such particles should be equal to each other, which is not the
case for $m_{T_{1}}$ and $m_{T_{2A}}$. In our view, additional studies are
necessary to solve problems existing in the QCD sum rule analyses of the
resonance $X_{0}$ in the diquark-antidiquark picture.

Interesting analysis of the ground-state and radially excited tetraquark $%
T_{2}$ was performed in Ref.\ \cite{He:2020jna}. In this paper the mass of
the particles $T_{2}(1S)$ and $T_{2}(2S)$ were found equal to $2360~\mathrm{%
MeV}$ and $2860~\mathrm{MeV}$, respectively. As a result, the resonance $%
X_{0}$ was interpreted there as the excited tetraquark $T_{2}(2S)$.

The enhancements in the $D^{-}K^{+}$ mass distribution labeled by $X_{0}$
and $X_{1}$ may have alternative origin \cite{Liu:2020orv}. Thus, the
authors of Ref.\ \cite{Liu:2020orv} investigated the process $%
B^{+}\rightarrow D^{+}D^{-}K^{+}$ via $\chi _{c1}D^{\ast -}K^{\ast +}$ and $%
D_{sJ}\overline{D}_{1}^{0}K^{0}$ rescattering diagrams. It was argued that,
two resonance-like peaks obtained around thresholds $D^{\ast -}K^{\ast +}$
and $\overline{D}_{1}^{0}K^{0}$ may simulate the states\ $X_{0}$ and $X_{1}$
without a necessity to introduce genuine four-quark mesons. The observed
experimental peaks were explained there by the triangle singularities in the
scattering amplitudes located in the vicinity of the physical boundary.

Experimental results obtained by the LHCb collaboration do not raise doubts
about existence of the resonance-like enhancements $X_{0}$ and $X_{1}$ in
the $D^{-}K^{+}$ mass distribution. These structures were already considered
as meson molecules, diquark-antidiquark systems, rescattering effects. In
other words, there are different and controversial interpretations of the
structures $X_{0}$ and $X_{1}$ in the literature. Additional theoretical
efforts seem are required to clarify nature of these states.

\end{document}